\documentclass[]{spie}  
\usepackage[]{graphicx}
\usepackage{amsmath}
\usepackage{enumerate}
\usepackage{color}
\usepackage{cite}
\usepackage{mathptmx}      
\usepackage{bm}

\title{Nonclassical effects in two-photon interference experiments: event-by-event simulations}

\author{K. Michielsen\supit{a,}\supit{b}, F. Jin\supit{a}, and H. De Raedt\supit{c}
\skiplinehalf
\supit{a}
Institute for Advanced Simulation, J\"ulich Supercomputing Centre,
Forschungszentrum J\"ulich, D-52425 J\"ulich, Germany
\\
\supit{b}
RWTH Aachen University, D-52056 Aachen, Germany
\\
\supit{c}
Department of Applied Physics,
Zernike Institute for Advanced Materials,
University of Groningen, Nijenborgh 4, NL-9747 AG Groningen, The Netherlands
\\
}

\authorinfo{
Further author information: (Send correspondence to K. Michielsen)\\
K. Michielsen: E-mail: K.Michielsen@fz-juelich.de \\
F. Jin: E-mail: F.Jin@fz-juelich.de \\
H. De Raedt : E-mail: h.a.de.raedt@rug.nl \\
}

\newcommand\MT{Maxwell's theory}
\newcommand\QT{quantum theory}

\newcommand{\url}[1]{{\rm #1}}

\begin{document}
  \maketitle

\begin{abstract}
It is shown that both the visibility ${\cal V} = 1/2$ predicted for two-photon
interference experiments with two independent sources\textcolor{black}{, like the Hanbury Brown-Twiss experiment,} and the visibility ${\cal V} = 1$
predicted for two-photon interference experiments with a parametric down-conversion source\textcolor{black}{, like the Ghosh-Mandel experiment,}
can be explained
\textcolor{black}{by a discrete event simulation. This simulation approach reproduces the statistical distributions of wave theory
not by requiring the knowledge of the solution of the wave equation of the whole system
but by generating detection events one-by-one according to an unknown distribution.}
There is thus no need to invoke quantum theory to explain the so-called nonclassical effects
in the interference of signal and idler photons in parametric down conversion.
Hence, a revision of the commonly accepted criterion of the nonclassical nature of light\textcolor{black}{, ${\cal V} > 1/2$,} is called for.
\end{abstract}


\keywords{Interference, Hanbury Brown-Twiss experiment, Ghosh-Mandel experiment, quantum theory, discrete-event simulation}

\section{Introduction}

In classical physics, interference appears as a result of the superposition of waves
emitted by different (virtual) sources.
If the wave propagation is described by a linear equation, the waves do not interact
and the observed interference must be the result of the interaction of the waves with material substance.
For instance, according to Maxwell's theory, electromagnetic waves propagating in vacuum
do not interact~\cite{ROYC10,ROYC10b} but
interference phenomena may be observed when different (partial) waves impinge on an array of detectors, such
as a photographic plate or CCD camera, where they interact with the electrons, atoms etc. of the material.
Observed for the first time in Young's two-slit experiment~\cite{YOUN02}, interference played an
important role in the general acceptance of the wave character of light.

An interesting variant of Young's double slit experiment
involves a very dim light source so that on average only one photon is emitted by the source at any time.
Inspired by Thomson's idea that light consists of indivisible units that are more widely separated
when the intensity of light is reduced,~\cite{THOM08}
in 1909 Taylor conducted an experiment with a light source varying in strength and illuminating a needle thereby demonstrating
that the diffraction pattern observed with a feeble light source (exposure time of three months) was as sharp
as the one obtained with an intense source and a shorter exposure time.~\cite{TAYL09}
In 1985, a double-slit experiment was performed with a low-pressure mercury lamp and neutral density filters
to realize a very low-light level.~\cite{TSUC85}
It was shown that at the start of the measurement bright dots appeared at
random positions on the detection screen and that after a couple of minutes an interference pattern appeared.
Demonstration versions of double-slit experiments illuminated by strongly attenuated lasers are
reported in Refs.~\citen{PARK71,WEIS03} and in figure 1 of Ref.~\citen{DIMI08}.
However, attenuated laser sources are imperfect single-photon sources.
Evidence for the indivisible, particle-like character comes from
experiments~\cite{GRAN86} in which these photons are sent to a beam splitter and the measured coincidence
of detector clicks on both output ports of the beam splitter is much smaller
than what might be expected on the basis of classical wave theory.
Light from sources attenuated to the single-photon level never antibunches,
which means that the anticorrelation parameter $\alpha\ge 1$.
For a real single-photon source $0<\alpha<1$.
In 2005, a variation of Young's experiment was performed with a Fresnel biprism and a single-photon source based on the pulsed, optically excited photoluminescence
of a single N-V colour centre in a diamond nanocrystal.~\cite{JACQ05}
In this two-beam experiment there is always only one photon between the source and the detection plane.
It was observed that the interference pattern gradually builds up starting from a couple of dots spread over the screen for small exposure times.
A time-resolved two-beam experiment has been reported in Refs.~\citen{SAVE02,GARC02}.
Recently, a temporally and spatially resolved two-beam experiment has been performed with entangled photons, providing
insight in the dynamics of the build-up process of the interference pattern.~\cite{KOLE13}

The common observation in interference experiments with single-particles, where
``single particle'' can be read as photon, electron~\cite{DONA73,MERL76,TONO89,HASS10,ROSA12}
neutron~\cite{ZEIL88,RAUC00}, atom~\cite{KEIT91,CARN91} and molecule~\cite{ARND99,BREZ02,JUFF12}
is that individual detection events gradually build up a pattern which we find to exhibit certain regularities
and that after collecting many detection events this pattern can be described by wave theory,
hence the name interference pattern.
In trying to give a pictorial, cause-and-effect, description of what is going on in these experiments, it is
commonly assumed that there is a one-to-one correspondence between an emission
event, ``the departure of a single particle from the source'' and a detection event,
``the arrival of the single particle at the detector''.
This assumption might be wrong.
The only definite fact is that there are detection events
and all other assumptions are inferences based on models that one has in mind.
In other words, we can only be certain about the fact that only one detector fired, not about
whether this event was caused by the ``indivisible photon''.

In general, an optical interference experiment entails several classical light sources
(not necessarily primary sources)
and several detectors which measure the resulting light intensity at various positions.
Adding equipment to accumulate the time average of the product of the detector signals
allows for the measurement of the second and higher order intensity correlations.
In quantum optics, the sources are replaced by single photon sources (the primary source commonly said to create
single photons or $N$-photon entangled states with $N\ge 2$) and single photon detectors. A coincidence circuit is added
to the experimental setup to measure (the absence of) coincidences in the photon counts.

In this paper we limit the number of sources and the number of detectors to two.
Interference is then characterized by the dependence of the resulting light intensity
or of the second order intensity correlations on certain phase shifts.
The Hanbury Brown-Twiss (HBT) effect was one of the first observations that
demonstrated interference in the intensity-intensity correlation functions~\cite{HANB56}.
HBT showed that under conditions for which the usual two-beam interference fringes measured by each of the two detectors vanish,
the correlated intensities of the two-detectors can still show interference fringes.
For two completely independent sources, be it classical light sources or single-photon sources,
the visibility of this second-order intensity interference has an upperbound of 1/2~\cite{MAND99}.
For primary sources producing correlated photon pairs, such as parametric down-converting sources,
the two sources in an HBT-type of experiment can no longer be considered to be independent.
In that case the two sources are considered to emit exactly one photon of the correlated pair simultaneously.
Such sources provide a $100\%$ visibility of the second-order intensity correlation, exceeding the $50\%$ limit which is
a commonly accepted criterion of nonclassicality~\cite{MAND99}.
The first experiment devoted to demonstrate nonclassical second-order intensity interference effects in the absence of first order intensity interference
is probably the Ghosh-Mandel two-photon interference experiment of 1987~\cite{GHOS87}.
However, the effect is not limited to photons.
Second-order intensity interference effects have also been observed in two-atom interference experiments~\cite{YASU96,OTTL05,SCHE05,JELT07}
in which an expanding cloud of cooled atoms acts as a source, multi-channel
plate(s) detect the arrival and position of a particle, and time-coincidence techniques
are employed to obtain the two-particle correlations.
Also in Hanbury Brown-Twiss type of experiments with electrons second-order intensity interference effects have been
observed~\cite{OLIV99,KIES02}.
Both the intensity interference in the two-slit experiment and the
second-order intensity interference in Hanbury Brown and Twiss-type of
experiments have been attributed to the dual wave-particle character.

In earlier work, see Refs.~\citen{MICH11a,RAED12a,RAED12b} for reviews,
we have demonstrated that
discrete-event simulations of experiments faithfully reproduce experimental results
involving interference and entanglement, both with photons and neutrons.
Among other things, this work shows that interference is not necessarily a signature of the presence of waves
of some kind but can also appear as the collective result of particles which at any time do not
directly interact with each other.
In general, the event-based approach deals with the fact that experiments yield definite results, such as
for example the individual detector clicks that build up an interference pattern.
We call these definite results ``events''.
Instead of trying to fit the existence of these events
in some formal, mathematical theory, the event-based approach changes the paradigm
by directly searching for the rules that transform events
into other events and, which by repeated application,
yield frequency distributions of events that
agree with those predicted by classical wave or quantum theory.
Obviously, such rules cannot be derived from quantum theory or,
as a matter of fact, of any theory that is probabilistic in nature
simply because these theories do not entail a procedure (= algorithm)
to produce events themselves.

In this paper, we demonstrate that the second-order intensity interference with visibility 1/2 in a HBT experiment with two
independent single photon sources and with visibility 1 in the Ghosh-Mandel experiment can be entirely explained in terms of
this event-based model, that is in terms of a locally causal, modular, adaptive, classical (non-Hamiltonian) dynamical system.
Hence, there is no need to invoke quantum theory to explain the observations and the commonly accepted criterion
of the nonclassical nature of light needs to be revised.

\begin{figure}[t]
\begin{center}
\includegraphics[width=8cm ]{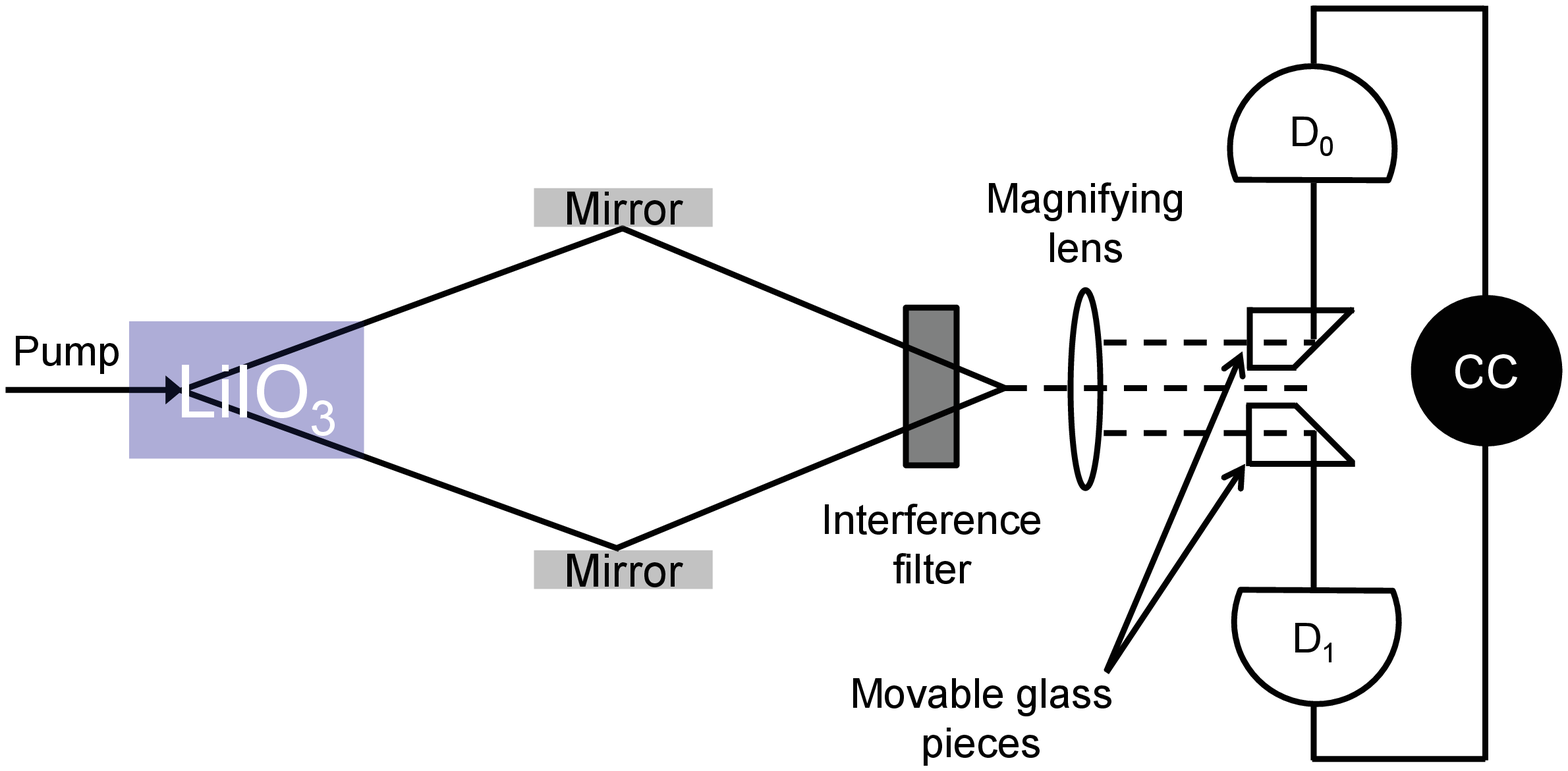}
\includegraphics[width=8cm]{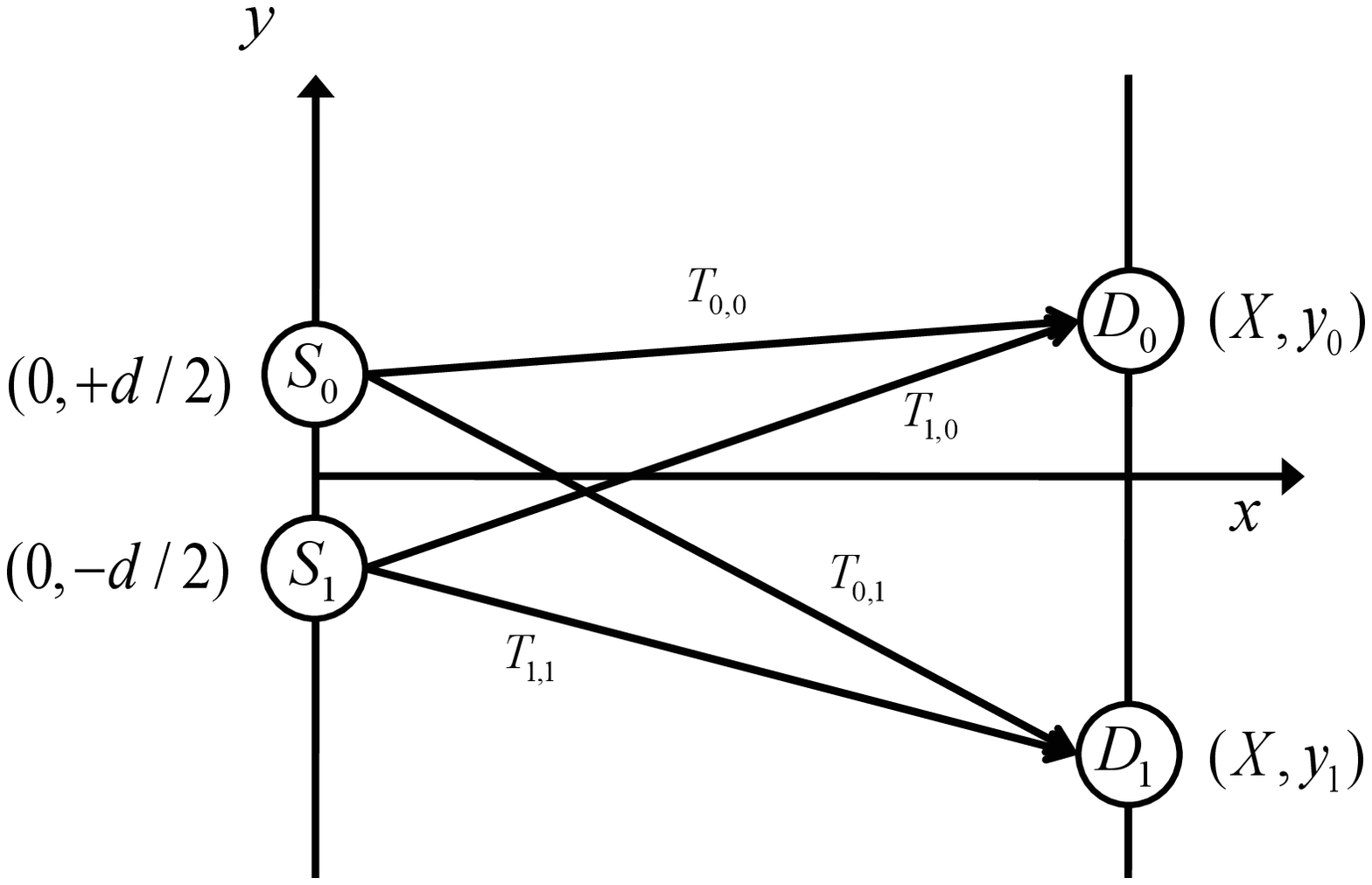}
\caption{%
Left: Diagram of the Ghosh--Mandel interference experiment~\cite{GHOS87}.
A source emits pairs of single-photons through spontaneous down-conversion
in a LiIO$_3$ crystal.
These photons leave the source in different directions.
Mirrors redirect the photons to the interference filter and a lens.
The two beams overlap at a distance of about $1\,$m from the crystal.
The resulting image is magnified by a lens and two movable glass pieces
are used to collect and redirect the photons to
the single-photon detectors $D_0$ and $D_1$,
the signals of which are fed into a coincidence counter CC.
Right: Schematic diagram of the Ghosh--Mandel experiment.
Single photons emitted from point sources $S_0$ and $S_1$ positioned at the $y$ axis and separated by a center-to-center
distance $d$ are registered by two detectors $D_0$ and $D_1$ positioned on a line at a distance $X$ from the $y$ axis.
The time of flight for each of the four possible paths from source $S_m$ to detector $D_n$
is denoted by $T_{m,n}$ where $m,n=0,1$.
}%
\label{fig.1}
\label{exphbt}
\end{center}
\end{figure}

\section{Second-order intensity interference}

In the context of the Ghosh--Mandel experiment, see Fig.~\ref{fig.1}(left), we may view the two mirrors as the two sources
that produce two overlapping beams of photons.
Hence, conceptually, this experiment can be simplified as shown in Fig.~\ref{exphbt}(right), which is
the schematic diagram of a HBT experiment~\cite{MAND99}.

A HBT experiment is nothing but a two-beam experiment with two sources and two detectors.
The two sources are positioned along the $y$-axis and are separated by a center-to-center distance $d$.
The two detectors are placed on a line at a distance $X$ from the $y$-axis.
Assume that source $S_m$ ($m=0,1$) emits coherent light of frequency $f$
and produces a wave with amplitude $A_m e^{i\phi_m}$ ($A_m$ and $\phi_m$ real).
For simplicity of presentation, we assume that $A_0=A_1=A$.
According to \MT, the total wave amplitude $B_n$ on detector $n$ is
\begin{equation}
B_n = A \left(e^{i(\phi_0+2\pi fT_{0,n})} + e^{i(\phi_1+2\pi fT_{1,n})}\right)
,
\end{equation}
where the time of flight for each of the four possible paths from source $S_m$ to detector $D_n$
is denoted by $T_{m,n}$ where $m,n=0,1$.
The light intensity $I_n = |B_n|^2$ on detector $D_n$ is given by
\begin{equation}
I_n = 2A^2\left\{ 1+ \cos \left[\phi_0-\phi_1+2\pi f(T_{0,n}-T_{1,n})\right]\right\}
.
\label{hbt0}
\end{equation}
If the phase difference $\phi_0-\phi_1$ in Eq.~(\ref{hbt0}) is fixed, the usual two-beam  (first-order) interference fringes are observed.

The essence of the HBT experiment is that if the phase difference $\phi_0-\phi_1$ is a random variable
(uniformly distributed over the interval $[0,2\pi[$) as a function of observation time,
these first-order interference fringes vanish because
\begin{equation}
\langle I_n\rangle = 2A^2,
\label{hbt1}
\end{equation}
where $\langle . \rangle$ denotes the average over the variables $\phi_0$ and $\phi_1$.
However, the average of the product of the intensities $I_0$ and $I_1$ is given by
\begin{equation}
\langle I_0I_1\rangle = 4A^4\left( 1+ \frac{1}{2}\cos 2\pi f \Delta T\right)
,
\label{hbt2}
\end{equation}
where $\Delta T =(T_{0,0}-T_{1,0})-(T_{0,1}-T_{1,1})$.
Accordingly, the intensity-intensity correlation Eq.~(\ref{hbt2}) exhibits second-order interference fringes,
a manifestation of the so-called HBT effect.
From Eqs.~(\ref{hbt1}) and (\ref{hbt2}), it follows that the visibility of the signal $I$, defined by
\begin{equation}
{\cal V}= \frac{\max(I)-\min(I)}{\max(I)+\min(I)}
,
\label{hbt2a}
\end{equation}
is given by ${\cal V}=0$ and ${\cal V}=1/2$ for the first-order and second-order intensity interference, respectively.

Treating the electromagnetic field as a collection of bosons changes Eq.~(\ref{hbt2}) into~\cite{MAND99}
\begin{equation}
\langle I_0I_1\rangle^{\mathrm{bosons}} = 4A^4\left( 1+ \cos 2\pi f \Delta T\right)
.
\label{hbt2b}
\end{equation}
Clearly, for bosons, the visibility of the second-order intensity interference is ${\cal V}=1$.

Considering the situation in which the two independent sources $S_0$ and $S_1$
are replaced by sources that emit simultaneously exactly one photon of a correlated photon pair
emitted by a parametric down-conversion source
gives a similar expression for the average of the product of the intensities $I_0$ and $I_1$ as given
by Eq.(~\ref{hbt2b}). Hence, also in this case ${\cal V}=1$ for the second-order intensity interference.

In the two-beam experiment interference appears in its most pure form because
the phenomenon of diffraction is absent.
If we assume that the detectors cannot communicate with each other,
that there is no direct communication between the particles involved
and that it is indeed true that individual pairs of particles build up the interference pattern one by one,
just looking at Fig.~\ref{exphbt} leads to the logically unescapable
conclusion that the interference can only be due to the internal operation of the detector~\cite{PFlE67}.
Detectors that simply count the incoming photons are not sufficient to explain the appearance of an interference pattern
and apart from the detectors there is nothing else that can cause the interference pattern to appear.
We now discuss an event-based model of a detector that can cope with this problem~\cite{MICH11a}.

\section{Simulation model}\label{sec2}

The model discussed in this paper builds on our earlier work~\cite{RAED05d,RAED05b,RAED05c,JIN10a,JIN10b,MICH11a}.
In short, in our simulation approach, a photon is viewed as a messenger that carries a message
and material is regarded as a message processor.
Evidently, the messenger itself can be thought of as a particle.
For the present purpose, it suffices to encode in the message, the time of flight of the particle.
The interaction of the photons with material translates into a processing unit receiving,
manipulating and sending out messages.
Note that we explicitly prohibit two particles from communicating directly
and that interference results from the processing of individual particles only~\cite{RAED05d,RAED05b,RAED05c,JIN10a,JIN10b,MICH11a}.

We now explicitly describe the model, that is
we specify the message carried by the messengers, the algorithm for simulating
a detector ( = processing unit), and the simulation procedure itself.

\textbf{Messenger:} The messenger can be regarded as a particle which travels with velocity $c$ in the direction $\mathbf{q}/q$.
Each messenger carries with it a harmonic oscillator which vibrates with frequency $f$.
It may be tempting to view the messenger with its message as a plane wave
with wave vector $\mathbf{q}$, the oscillator being one of the two electric field
components in the plane orthogonal to $\mathbf{q}$.
However, this analogy is superfluous and should not be stretched too far.
As there is no communication/interaction between the messengers
there is no wave equation (i.e.~no partial differential equation)
that enforces a relation between the messages carried by different messengers.
Indeed, the oscillator carried by a messenger never interacts with the oscillator
of another messenger, hence the motion of these pairs of oscillators is not governed by a wave equation.
Naively, one might imagine the oscillators tracing out a wavy pattern as they travel through space.
However, as there is no relation between the times at which the messengers leave the source,
it is impossible to characterize all these traces by a field that
depends on one set of space-time coordinates, as required for a wave theory.
It is convenient (though not essential) to represent the
message, that is the oscillator,
by a two-dimensional unit vector ${\mathbf y}=\left( \cos \psi,\sin \psi\right) $ where $\psi=2\pi f t+\delta$.
Here, $t$ is the time of flight of the particle and $\delta$ is a phase shift.
Pictorially, the message is nothing but a representation of the hand of a clock which
rotates with period $1/f$ and is running ahead by a time related to the phase $\delta$.
A processing unit has access to this data and may use the messenger's internal clock
to determine how long it took for the messenger to reach the unit.

\textbf{Source:} A source creates a messenger 
with its phase $\delta$ set to some randomly chosen value.
Initially its time of flight $t$ is zero as it  is determined by the arrival of the messenger at a processing unit.
A pseudo-random number determines to which of the two detectors the messenger travels.

\textbf{Single-photon detector:} In reality, photon detection is the result of a complicated
interplay of different physical processes~\cite{HADF09}.
In essence, a light detector consists of material that absorbs light.
The electric charges that result from the absorption process are then amplified,
chemically in the case of a photographic plate or electronically in the case of photodiodes or photomultipliers.
In the case of photomultipliers or photodiodes,
once a photon has been absorbed (and its energy ``dissipated'' in the detector material)
an amplification mechanism (which requires external power/energy) generates
an electric current (provided by an external current source)~\cite{GARR09,HADF09}.
The resulting signal is compared with a threshold that is set
by the experimenter and the photon is said to have been detected
if the signal exceeds this threshold~\cite{GARR09,HADF09}.
In the case of photographic plates, the chemical process that occurs when photons
are absorbed and the subsequent chemical reactions that
renders visible the image serve similar purposes.
Photon detectors, such as a photographic plate of CCD arrays, consist of many identical detection units
each having a predefined spatial window in which they can detect photons.
In what follows, each of these identical detection units will be referred to as a detector.
By construction, these detector units operate completely independently from and also do not communicate with each other.

An event-based model for the detector cannot be ``derived'' from  \QT\
simply because \QT\ has nothing to say about individual events
but predicts the frequencies of their observation only~\cite{HOME97}.
Therefore, any model for the detector that operates on the level of single events
must necessarily appear as ``ad hoc'' from the viewpoint of  \QT.
The event-based detector model that we employ in this paper should not be regarded
as a realistic model for say, a photomultiplier or a photographic plate and the
chemical process that renders the image.
In the spirit of Occam's razor, the very simple event-based model captures the salient features
of ideal (i.e. 100\% efficient) single-photon detectors.

The key element of the event-by-event
approach is a processing unit that is adaptive, that is it can learn
from the messengers that arrive at its input ports~\cite{RAED05d,RAED05b,MICH11a}.
The event-based detection unit
consists of an input stage called deterministic learning machine (DLM)~\cite{RAED05d,RAED05b},
a transformation stage, and an output stage.
The processing unit should act as a detector for individual messengers which may come
from several different directions. Therefore,
this processing unit
has $K$ input ports, a parameter that allows the machine to resolve $K$ different directions.
In what follows, we give the description of the detector model for general $K$ but to simulate
the Ghosh-Mandel experiment, it is sufficient to consider the case $K=2$.

\textit{Input stage:} Representing the arrival of a messenger at port $1\le k\le K$ by
the vector ${\bf v}=(v_1,\ldots,v_K)^T$ with $v_i=\delta_{i,k}$ ($i=1,\ldots K$)
the internal vector is updated according to the rule
\begin{equation}
\mathbf{x} \leftarrow \gamma \mathbf{x} + (1-\gamma) \mathbf{v}
,
\label{det0}
\end{equation}
where $\mathbf{x}=(x_1,\ldots,x_{K})^T$, $\sum_{k=1}^K x_k=1$, and $0\le\gamma<1$.
The elements of the incoming message $\mathbf{y}$ are written in internal register $\mathbf{Y}_k$
\begin{equation}
\mathbf{Y}_k\leftarrow\mathbf{y}
,
\label{det1}
\end{equation}
while all the other $\mathbf{Y}_{i}$ ($i\not=k$) registers remain unchanged.
Thus, each time a messenger arrives at one of the input ports, say $k$,
the DLM updates all the elements of the internal vector $\mathbf{x}$,
overwrites the data in the register $\mathbf{Y}_{k}$
while the content of all other $\mathbf{Y}$ registers remains the same.

\textit{Transformation stage}: The output message generated by the transformation stage is
\begin{equation}
\mathbf{T}=\mathbf{x}\cdot\mathbf{Y}=\sum_{k=1}^K x_k \mathbf{Y}_k
,
\label{det2}
\end{equation}
which is a two-component vector. Note that $|\mathbf{T}|\le1$.

\textit{Output stage}: As in all previous event-based models for the single-photon detectors,
the output stage generates a binary output signal $z=0,1$
but the output message does not represent a photon:
It represents a ``no click'' or ``click'' if $z=0$ or $z=1$, respectively.
To implement this functionality, we define
\begin{equation}
z = \Theta(|\mathbf{T}|^2- {\cal R})
,
\label{det3}
\end{equation}
where $\Theta(.)$ is the unit step function and $0\leq {\cal R} <1$
are uniform pseudo-random numbers (which are different for each event).
The parameter $0\le \gamma<1$ can be used to control the operational mode of the unit.
From Eq.~(\ref{det3}) it follows that the frequency of $z=1$ events depends
on the length of the internal vector $\mathbf{T}$.

Note that in contrast to experiment, in a simulation, we could register both the $z=0$ and $z=1$ events.
Then the sum of the $z=0$ and $z=1$ events is equal to the number of input messages.
In real experiments, only $z=1$ events are taken as evidence that a photon has been detected.
Therefore, we define the total detector count by
\begin{equation}
N_{\mathrm{count}}=\sum^{N}_{l=1}z_l,
\label{det4}
\end{equation}
where $N$ is the number of messages received and $l$ labels the events.
In other words, $N_{\mathrm{count}}$ is the total number of one's generated by the detector unit.

Comparing the number of ad hoc assumptions and unknown functions that enter quantum theoretical treatments
of photon detectors~\cite{GARR09} with the two parameters $\gamma$ and $K$ of the event-based detector model,
the latter has the virtue of being extremely simple while providing a description of
the detection process at the level of detail, the single events, which in any case is outside the scope of \QT.

\textbf{Simulation procedure:} Before the simulation starts we set ${\bf x}=(1,0,\ldots ,0)^T$
and we use pseudo-random numbers ${\cal R}$ to set $\mathbf{Y}_{k}=(\cos2\pi{\cal R},\sin2\pi{\cal R})$ for $k=1,\ldots,K$.
Next, we generate $N_{\mathrm{tot}}$ pairs of messengers (i.e. $2N_{\mathrm{tot}}$ messengers),
send them to the detectors, determine the detector count $N_{\mathrm{count}}$ at $D_0$ and $D_1$
and count the coincidences. In the simulation always two messengers travel to the detectors, one generated at source $S_0$ and one at
source $S_1$. Hence, once a pair of messengers is generated a detector can generate no click, one click or two clicks.
Only when both detectors generate a click the coincidence count $N_{\mathrm {coincidence}}$ is enhanced by one.

\section{Simulation results}

\subsection{Detection efficiency}

The efficiency of the detector model is determined by simulating an experiment
that measures the detector efficiency, which for a single-photon detector is defined
as the overall probability of registering a count if a photon arrives at the detector~\cite{HADF09}.
In such an experiment a point source emitting single particles is placed far away from a single detector.
As all particles that reach the detector have the same time of flight (to a very good approximation), all the
particles that arrive at the detector will carry nearly the same message $\mathbf{y}$ which is encoding the time of flight.
Furthermore, they arrive at the same input port, say $q$.
As a result ${\mathbf x}$ (see Eq.~(\ref{det0})) rapidly converges to the vector with
$x_i\rightarrow \delta_{i,q}$ and, as $\mathbf{y}$ is a unit vector, we have $|\mathbf{T}|\approx 1$,
implying that the detector clicks almost every time a photon arrives.
Thus, for our detector model, the detection efficiency as defined for real detectors~\cite{HADF09} is very close
to 100\% (results not shown).

\subsection{Hanbury Brown-Twiss experiment}\label{HBT}

\begin{figure}[t]
\begin{center}
\includegraphics[width=8cm]{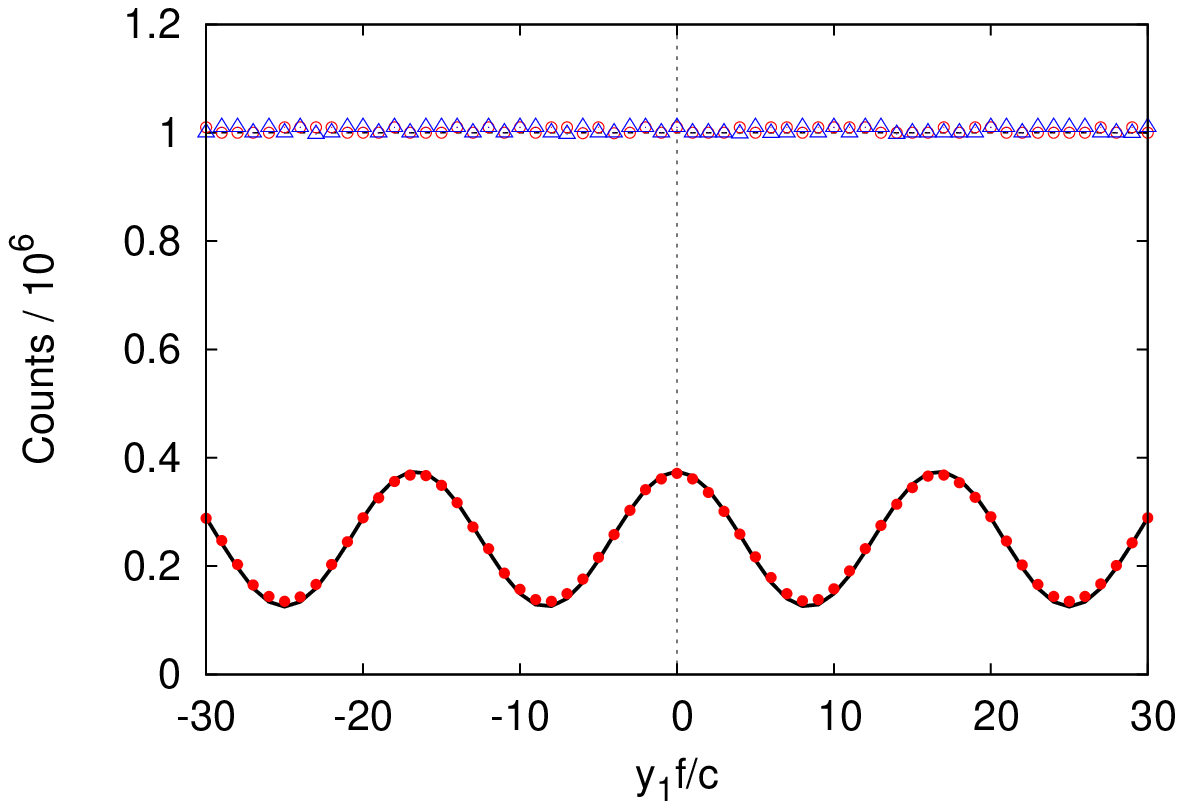}
\includegraphics[width=8cm]{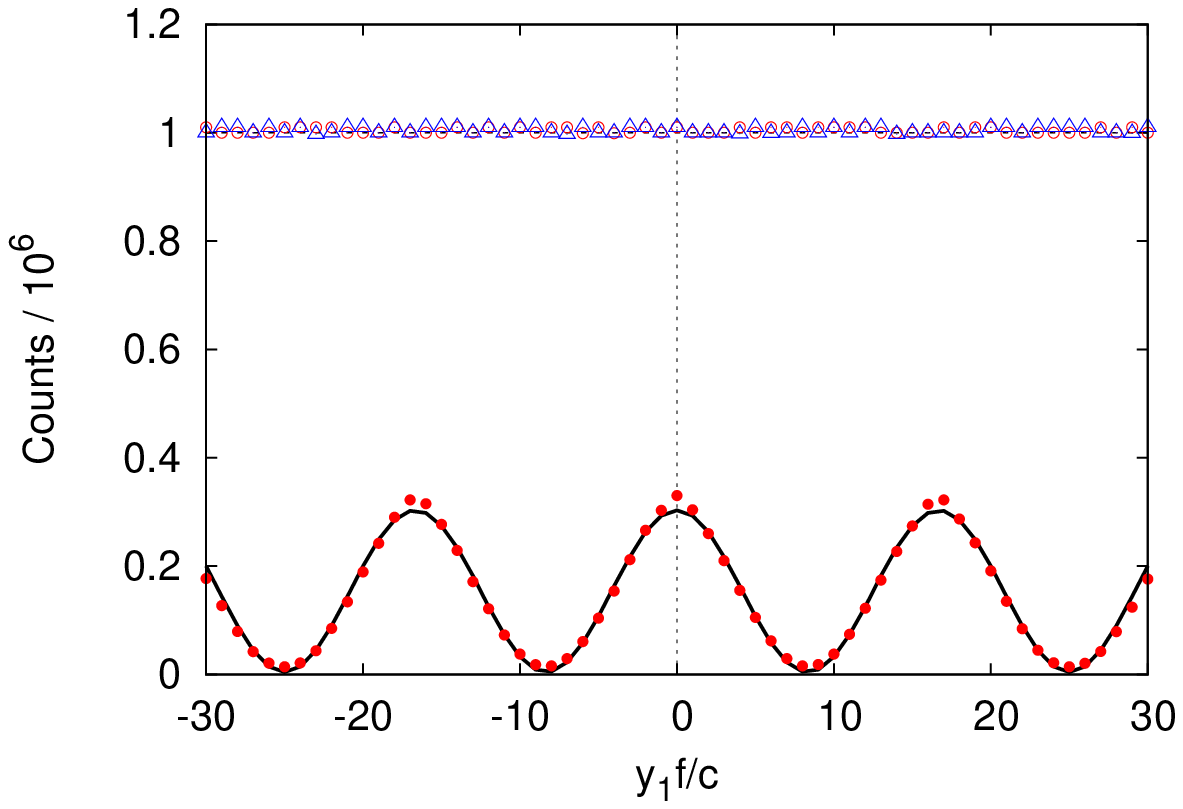}
\caption{%
Left:
Simulation data of the single-particle and two-particle counts
for the HBT experiment depicted in Fig.~\ref{exphbt}.
Red open circles (blue open triangles): results for the counts $N_{\mathrm {count}}$ of detector $D_0$ ($D_1$),
showing that there is no first-order intensity interference.
Red closed circles: results for the coincidence counts $N_{\mathrm{coincidence}}$.
The dashed and solid lines represent the theoretical predictions $N_{tot}/2$
and Eq.~(\ref{hbt4}) for the single detector and coincidence counts, respectively.
Simulation parameters: $N_{\mathrm{tot}}=2\times 10^6$ events per $y_1f/c$-value,
$N_{\mathrm{F}}=50$, $X=100000c/f$, $d=2000c/f$, $\gamma=0.99$ and $K=2$.
Right: Same as left except that time-delays are accounted for
through the model defined by Eq.~(\ref{hbt6}).
Simulation parameters: $T_{\mathrm{max}}/f=1000$, $W/f=1$, $h=8$.
The dashed and solid lines are least-square fits
to $a'_2N_{tot}$ and $a'_4N_{tot}(1+b'_4\cos2\pi f\Delta t)$
for the single detector and coincidence counts, $N_{\mathrm{count}}$ and $N_{\mathrm{coincidence}}$, respectively.
The values of the fitting parameters are
$a'_2=0.502$, $a'_4=0.077$ and $b'_4=0.974$.
}
\label{simhbt0}
\label{simhbt1}
\end{center}
\end{figure}

In Fig.~\ref{simhbt0} we present the simulation results for the HBT experiment depicted in Fig.~\ref{exphbt}(right).
For simplicity, we have put detector $D_0$ at $(X,0)$ and plot the single detector and
coincidence counts as a function of the $y$-position of detector $D_1$.
In each simulation step, both sources $S_0$ and $S_1$ create a messenger 
with some randomly chosen phase being the only initial content of the messages $\mathbf{y}_{m}$ ($m=0,1$).
The phases are kept fixed for $N_{\mathrm{F}}$ successive pairs of messengers.
The total number of emitted pairs is denoted by $N_{\mathrm {tot}}$.
Two pseudo-random numbers are used to determine whether the messengers travel to detector $D_0$ or $D_1$.
The time of flight for the messenger travelling from source $S_m$ to detector $D_n$ is given by
\begin{equation}
T_{m,n}=\frac{\sqrt{X^2+((1-2m)d/2 - y_n)^2}}{c}
,
\label{hbt4a}
\end{equation}
where $m,n=0,1$. The time of flight $T_{m,n}$ is added to the message $\mathbf{y}_{m}$ before the message is processed by the
corresponding detector $D_n$. The messages are the only input to the event-based model.
As Fig.~\ref{simhbt0}(left) shows, averaging over the randomness in the initial messages (random phases) wipes
out all interference fringes in the single-detector counts, in agreement with \MT.
We find that the number of single-detector counts $N_{\mathrm {count}}$ fluctuates around $N_{\mathrm{tot}}/2$, as expected from wave theory.
Similarly, the data for the coincidence counts are in excellent agreement with the theoretical prediction for
the simulation model
\begin{equation}
N_{\mathrm{coincidence}} = \frac{N_{\mathrm{tot}}}{8} \left(1+ \frac{1}{2}\cos 2\pi f \Delta T\right)
,
\label{hbt4}
\end{equation}
and, disregarding the prefactor $N_{\mathrm{tot}}/8$, also in qualitative agreement with the predictions of wave theory.

For simplicity, we have confined the above presentation to the case of a definite polarization.
Simulations with randomly varying polarization (results not shown) are also in concert
with \MT.

\subsection{Ghosh-Mandel experiment}

From Eq.~(\ref{hbt4}), it follows that the visibility of the interference fringes, defined by
\begin{equation}
{\cal V}= \frac{\max(N_{\mathrm{coincidence}})-\min(N_{\mathrm{coincidence}})}{\max(N_{\mathrm{coincidence}})+\min(N_{\mathrm{coincidence}})}
,
\label{hbt5}
\end{equation}
cannot exceed 1/2.
It seems commonly accepted that the visibility of a two-photon interference experiment exceeding 1/2
is a signature of the nonclassical nature of light.

As two-photon interference experiments, such as the Gosh-Mandel experiment~\cite{GHOS87},
employ time-coincidence to measure the intensity-intensity correlations,
it is quite natural to expect that a model that purports to explain the observations accounts for the time delay
that occurs between the time at which a particle arrives at a detector and the actual click of that detector.
In \QT, time is not an observable and can therefore not be computed within the theory proper.
Hence there is no way that these time delays, which are being measured, can be accounted for by  \QT.
Consequently, any phenomenon that depends on these time delays must find an explanation
outside the realm of  \QT\ (as it is formulated to date).

It is straightforward to add a time-delay mechanism to the event-based model of the detector.
For simplicity, let us assume that the time delay for the detector click is given by
\begin{equation}
t_{\mathrm{delay}}=T_{m,n}-T_{\mathrm{max}}(1-|\mathbf{T}|^2)^h\ln{\cal R}
,
\label{hbt6}
\end{equation}
where $0<{\cal R}<1$ is a pseudo-random number, and $\mathbf{T}$ is given by Eq.~(\ref{det2}).
The time scale $T_{\mathrm{max}}$ and the exponent $h$ are
free parameters of the time-delay model.
Note that $t_{\mathrm{delay}}-T_{m,n}$ is a pseudo-random variable
drawn from an exponential distribution with mean $T_{\mathrm{max}}(1-|\mathbf{T}|^2)^h$.
Coincidences are counted by comparing the difference between the delay times
of detectors $D_0$ and $D_1$ with a time window $W$.

From the simulation results presented in Fig.~\ref{simhbt1}(right),
it is clear that by taking into account that there are fluctuations
in the time delay that depend on the time of flight and the internal state
of the detector, the visibility changes from ${\cal V}=1/2$ to ${\cal V}\approx1$.
The simulation data is represented (very) well by $N'_{\mathrm{count}}\approx N_{\mathrm{tot}}/2$ and
\begin{equation}
N'_{\mathrm{coincidence}} \approx a'_4 N_{tot}\left(1+ \cos 2\pi f \Delta T\right)
,
\label{hbt10}
\end{equation}
where the prime indicates that the model incorporates the time-delay mechanism
and $a^\prime_4$ is a fitting parameter which depends on the details
of the time-delay mechanism.
As expected, the use of a narrow time window leads to a significant reduction (by a factor $a'_4=0.077$) of the total coincidence count.
These results demonstrates that a purely classical corpuscular model of a two-photon interference experiment
can yield visibilities that are close to one.
Hence, the commonly accepted criterion of the nonclassical nature of light needs to be revised.

The time delay model Eq.~(\ref{hbt6}) is perhaps one of the simplest that yield interesting results but
it is by no means unique and can only be scrutinized on the basis of accurate experimental data
which, unfortunately, do not seem to be available thus far.

\begin{figure}[t]
\begin{center}
\includegraphics[width=12cm]{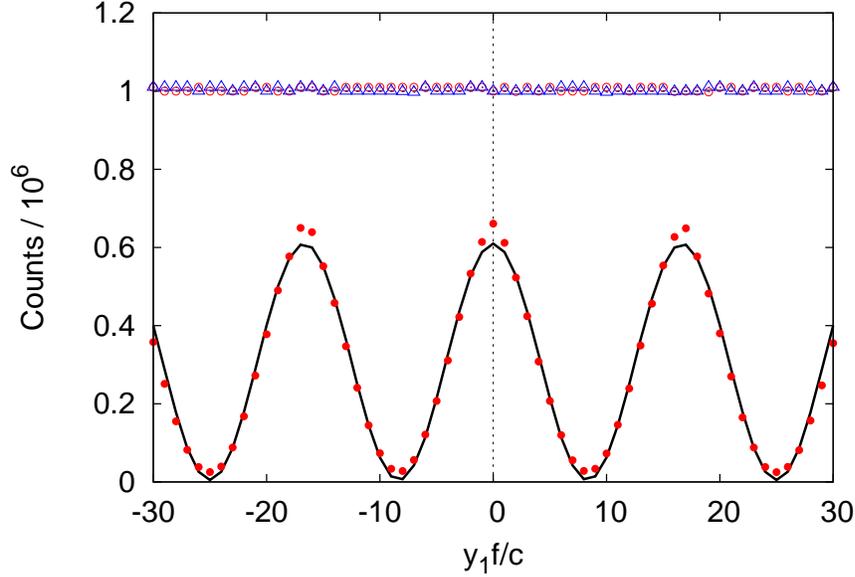}
\caption{%
Same as Fig.~\ref{simhbt1} except that the two sources never send their
particles to the same detector, mimicking bosons (see text).
The values of the fitting parameters are $a''_2=0.502$, $a''_4=0.154$ and $b''_4=0.985$.
}
\label{simhbt2}
\end{center}
\end{figure}

\subsection{Bosons}

If we exclude the possibility that the two sources send their particles
to the same detector, the event-based approach produces results
that are reminiscent of the quantum theoretical description in terms of bosons.
In Fig.~\ref{simhbt2}, we present the results of such a simulation,
using the same model parameters as those used to produce the results
of Fig.~\ref{simhbt1}.
From Figs.~\ref{simhbt1} and \ref{simhbt2}, it is clear
that the maximum amplitude of the two-particle interference signal
of the latter is two times larger than that of the former (the ``classical'' case),
as expected for bosons.
The simulation data is represented (very) well by $N''_{\mathrm{count}}\approx N_{\mathrm{tot}}/2$
\begin{equation}
N''_{\mathrm{coincidence}} \approx a''_4 N_{tot} \left(1+ \cos 2\pi f \Delta T\right)
,
\label{hbt11}
\end{equation}
where the double prime indicates that the model incorporates the time-delay mechanism
and that the possibility that the two sources send their particles
to the same detector has been excluded.

\subsection{Non-monochromatic sources}

All the results presented above have been obtained by assuming that the
beams of particles are strictly monochromatic, meaning that the frequency
$f$ of the oscillators carried by the particles is fixed.
A more realistic simulation of
the pairs of photons created by the parametric down-conversion process
requires that the frequencies $f_1$ and $f_2$ of the messages carried by the pair of particles
satisfy energy conservation, meaning that $f_1+f_2=f_0$ where $f_0$ is the frequency
of the pump beam~\cite{RUBI92,SHIH93,SHIH94,GARR09}.
It is straightforward to draw the
frequencies $f_1$ (and therefore $f_2=f_0-f_1$) from a specified
distribution, such as Gaussian or a Lorentzian~\cite{RUBI92,GARR09}.
In the simulation, each created particle pair would then correspond
to one message characterized by a frequency $f_1$
and another one by frequency $f_2$.
The detectors simply sum all the contributions (taking into account
the differences in the factors $f_m T_{m,n}$),  just as in the wave mechanical picture.

\section{Conclusion}

We have shown that the so-called nonclassical effects observed in two-photon interference experiments
with a parametric down-conversion source can be explained in terms of
\textcolor{black}{an event-based simulation approach.
The statistical distributions which are observed in these two-photon interference experiments and which are usually thought to be of quantum mechanical origin,
are shown to emerge from a time series of discrete events generated by causal adaptive systems, which in principle could be built
using macroscopic mechanical parts.}
The high visibility, ${\cal V}=1$, in this type of experiment which is commonly considered
as a signature of two-photon light, \textcolor{black}{is} in contrast
to the visibility ${\cal V}=1/2$ obtained in the same experiment with a classical light source.
According to Ref.~\citen{AGAF08}, the existence of high-visibility interference
in the third and higher orders in the intensity cannot be considered as a signature of
three- or four-photon interference, because
high-visibility interference is also observed in Hanbury Brown-Twiss type interference experiments with classical light.
Hence, although the case of second-order intensity interference seemed to be different from the higher orders,
our work demonstrates that also for the second order intensity interference the value of the visibility
cannot be used as a measure for the ``quantum nature'' of the source.
Elsewhere, we have shown that third-order intensity interference in a Hanbury Brown-Twiss type of experiment with two
sources emitting uncorrelated single photons can be modeled by an event-based model as well~\cite{JIN10a}.
Both the interference experiment with a classical light source as well as the interference experiment with the
parametric down-conversion source can therefore be explained entirely in terms of a discrete-event simulation
model which does not rely on wave theory and
in which the clicks of the detectors are interpreted as the arrival of ``particles''.

\section{Acknowledgment}
This work is partially supported by NCF, the Netherlands.

\bibliographystyle{spiebib}   
\bibliography{/d/papers/all13}   

\end{document}